\documentclass[american,aip,jcp,notitlepage,reprint]{revtex4-1}
\usepackage[T1]{fontenc}
\usepackage[latin9]{inputenc}
\usepackage{array}
\usepackage{float}
\usepackage{multirow}
\usepackage{amsmath}
\usepackage{amssymb}
\usepackage{graphicx}

\makeatletter

\providecommand{\tabularnewline}{\\}

\makeatother

\usepackage{babel}
\begin{document}

\title{Equilibrium configurations of large nanostructures using the embedded
saturated-fragments stochastic density functional theory}

\author{Eitam Arnon}

\affiliation{Fritz Haber Center for Molecular Dynamics, Institute of Chemistry,
The Hebrew University of Jerusalem, Jerusalem 91904, Israel}

\author{Eran Rabani}
\email{eran.rabani@berkeley.edu}

\affiliation{Department of Chemistry, University of California and Materials Science
Division, Lawrence Berkeley National Laboratory, Berkeley, California
94720, U.S.A.}

\affiliation{The Raymond and Beverly Sackler Center for Computational Molecular
and Materials Science, Tel Aviv University, Tel Aviv, Israel 69978}

\author{Daniel Neuhauser}
\email{dxn@chem.ucla.edu}

\affiliation{Department of Chemistry, University of California at Los Angeles,
CA-90095 USA}

\author{Roi Baer}
\email{roi.baer@huji.ac.il}

\affiliation{Fritz Haber Center for Molecular Dynamics, Institute of Chemistry,
The Hebrew University of Jerusalem, Jerusalem 91904, Israel}
\begin{abstract}
An \emph{ab initio} Langevin dynamics approach is developed based
on stochastic density functional theory (sDFT) within a new \emph{embedded
saturated } \emph{fragment }formalism, applicable to covalently bonded
systems. The forces on the nuclei generated by sDFT contain a random
component natural to Langevin dynamics and its standard deviation
is used to estimate the friction term on each atom by satisfying the
fluctuation\textendash dissipation relation. The overall approach
scales linearly with system size even if the density matrix is not
local and is thus applicable to ordered as well as disordered extended
systems. We implement the approach for a series of silicon nanocrystals
(NCs) of varying size with a diameter of up to $3$nm corresponding
to $N_{e}=3000$ electrons and generate a set of configurations that
are distributed canonically at a fixed temperature, ranging from cryogenic
to room temperature. We also analyze the structure properties of the
NCs and discuss the reconstruction of the surface geometry.
\end{abstract}
\maketitle

\section{Introduction}

\emph{Ab initio }molecular dynamics based on density functional theory
(DFT) is becoming an important tool for studying the plethora of structural
and dynamical processes in a broad range of systems in material science,
chemistry, biology and physics.\cite{Car1985,Barnett1991,Payne1992,Kresse1993,Tuckerman1995,Marx2000,schlegel2001ab,Herbert2005,Raty2005,bockstedte1997density,cawkwell2015extended}
The application of this approach to very large systems is still limited
by the computational scaling of the electronic structure portion of
the calculation, regardless of whether one uses a Lagrangian-based
or Born-Oppenheimer-based methods. This is because of the cubic scaling
involved in solving the Kohn-Sham equations coupled with the need
to iterate to self-consistency or to propagate the Kohn-Sham (KS)
orbitals, as both of these options further increases the computational
times by an order of magnitude. 

Significant advances in these respects have been made along two major
directions. One primary direction is based on a Lagrangian formulation
of density functional theory~\cite{Car1985,cawkwell2015extended}
and circumvents the need for SCF iterations by propagation of the
KS orbitals. This venue does not eliminate the cubic scaling and is
therefore limited to relatively small systems. Another approach is
based on linear-scaling techniques~\cite{Goedecker1999,Gillan2007,Baer1997b,Soler2002},
that reduces the algorithmic complexity by finding the density matrix
directly, relying on its asymptotic sparseness in real-space. However,
sparsity sets in only for very large systems, limiting the applicability
sparse-matrix methods, especially in 3D.

In a recent set of papers we have introduced the stochastic DFT (sDFT)
methods~\cite{Baer2013,Neuhauser2014a,Gao2015,Neuhauser2015} which
scales linearly (or even sublinearly) with the system size and does
not rely on the sparsity of the density matrix. sDFT is a general
approach to electronic structure based on a stochastic process and
is applicable to extended ordered as well as disordered materials.
Some of the techniques we use, based on the stochastic trace formula.\cite{Hutchinson1990},
have been developed for tight-binding electronic structure \cite{Drabold1993,Wang1994b,Roeder1997},
for molecular electronics \cite{Baer2004c} and for multi-exciton
generation in nanocrsytals \cite{Baer2012a}. The success of sDFT
in reducing the scaling comes at a price of introducing a stochastic
error in all its predictions, including forces, and that precludes
application to \emph{ab initio }molecular dynamics. 

In this paper we show that sDFT can be used to study equilibrium structural
properties of large NCs, despite the statistical fluctuations in the
force estimates. For this, we invoke the Langevin equation following
the work of Attaccalite and Sorella,\cite{Attaccalite2008} and generate
a sequence of configurations distributed according to the canonical
ensemble. These configurations can be used in a variety of applications
for studying the structural, electronic and optical properties of
NCs. Here we demonstrate their use for studying the structural properties
of silicon nanocrystals (NCs) with a diameter of up to $3$~nm, and
$N_{e}=3000$ electrons. 

The paper includes development of the embedded saturated fragments
method which allows reducing the statistical errors in sDFT. This
new method is inspired by, but more general than, the embedded fragments
method developed in Ref.~\onlinecite{Neuhauser2014a}. It uses small
saturated fragments of the system, and carves out the relevant part
of the density to be embedded in the system. Hence, it is applicable
not only to clusters of molecules, like Ref.~\onlinecite{Neuhauser2014a},
but also to covalently-bonded systems such as silicon NCs. The method
is described in detail in Appendix~\ref{sec:The-embedded-fragments}. 

\section{\label{sec:Methods}Methods}

\subsection{Stochastic DFT}

Kohn-Sham density functional theory~\cite{Hohenberg1964,Kohn1965}
maps a system of $N_{e}$ interacting electrons in an external electron-nucleus
potential $v_{eN}\left(\mathbf{r}\right)=-\frac{e}{4\pi\epsilon_{0}}\sum_{N}\frac{Z_{N}e}{\left|\mathbf{r}-\mathbf{R}_{N}\right|}$,
where $\mathbf{R}_{N}$ ($N=1,2,...$) are the nuclei positions and
$Z_{N}e$ are their charge ($e$ is the electron charge), onto a system
of \emph{non-interacting }electrons (the KS system), having the same
ground-state density $n\left(\boldsymbol{r}\right)$. This mapping
is performed by solving the KS equations~\cite{Hohenberg1964,Kohn1965}
\begin{equation}
\hat{h}_{\text{KS}}\phi_{n}\left(\boldsymbol{r}\right)=\varepsilon_{n}\phi_{n}\left(\boldsymbol{r}\right),\label{eq:KS-equations}
\end{equation}
where the KS Hamiltonian is:

\begin{equation}
\hat{h}_{\text{KS}}=-\frac{\hbar^{2}}{2m_{e}}\nabla^{2}+v_{KS}\left(\mathbf{r}\right),\label{eq:hKS}
\end{equation}
and the KS potential $v_{KS}\left(\mathbf{r}\right)$ is the sum of
the external electron-nuclear potential $v_{eN}\left(\mathbf{r}\right)$,
the density-dependent Hartree potential $v_{H}\left(\mathbf{r}\right)=\frac{e^{2}}{4\pi\epsilon_{0}}\int\frac{n\left(\mathbf{r}'\right)}{\left|\mathbf{r-r'}\right|}d\mathbf{r}'$,
and the exchange-correlation potential $v_{xc}\left(\mathbf{r}\right)$:
\begin{equation}
v_{KS}\left(\mathbf{r}\right)=v_{eN}\left(\mathbf{r}\right)+v_{H}\left(\mathbf{r}\right)+v_{xc}\left(\mathbf{r}\right).\label{eq:KS-pot}
\end{equation}
In the KS system, the density is expressed in terms of the normalized
single electron KS eigenstates $\phi_{n}\left(\boldsymbol{r}\right)$
and eigenvalues $\varepsilon_{n}$ 
\begin{equation}
n\left(\boldsymbol{r}\right)=2\sum_{n}\theta\left(\mu-\varepsilon_{n}\right)\left|\phi_{n}\left(\boldsymbol{r}\right)\right|^{2},\label{eq:n(r)}
\end{equation}
where $\theta\left(x\right)$ is the Heaviside function and $\mu$
is the chemical potential chosen so that $2\sum_{n}\theta\left(\mu-\varepsilon_{n}\right)=N_{e}$.
Eqs.~(\ref{eq:KS-equations})-(\ref{eq:n(r)}) must be solved self-consistently,
since $\hat{h}_{\text{KS}}$ depends on the density. While the entire
scheme is a significant simplification over the original many-electron
problem, it remains a challenge for large systems since the computational
effort scales as $O\left(N_{e}^{3}\right)$. 

\begin{figure*}
\includegraphics[width=0.8\textwidth]{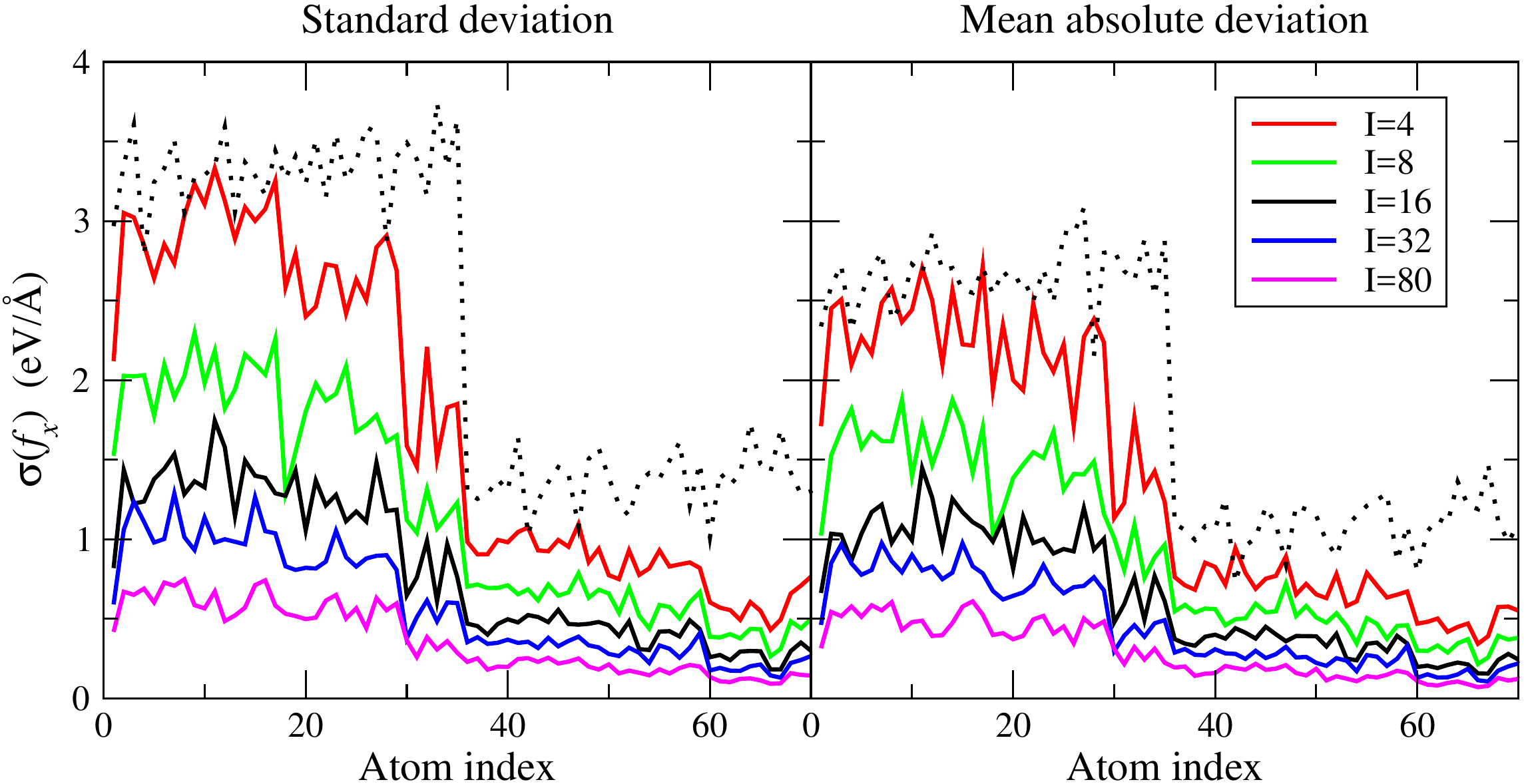}\caption{\label{fig:std of frags} The x-component of the atomic force statistics
for the $71$ atoms of $\text{Si}{}_{35}\text{H}_{36}$ calculated
by sDFT (black dashed line for $I=16$) and efsDFT (solid lines, using
passivated fragments of size smaller or equal to $\text{Si}{}_{5}$,
depending on the way surface atoms are treated). For each atom $\alpha=1,\dots,N$
and number of stochastic orbitals $I$ we present the standard deviation
(STD) $\sigma\left(f_{x}\right)=\sqrt{\left\langle \left(f_{\alpha}^{x}-\left\langle f_{\alpha}^{x}\right\rangle \right)^{2}\right\rangle _{I}}$
(left) and the mean-absolute-deviation (MAD) from the deterministic
DFT value, $\left\langle \left|f_{\alpha}^{x}-\left(f_{\alpha}^{x}\right)^{\text{det}}\right|\right\rangle _{I}$(right)
calculated using 60 independent efsDFT/sDFT runs. Atoms are ordered
by their distance from the origin, the first $35$ atoms are Si atoms
followed by $36$ H atoms. }
\end{figure*}

An important step towards reducing the computational scaling of KS-DFT
was recently proposed by Baer\emph{,} Neuhauser, and Rabani (BNR),\cite{Baer2013}
where the density of Eq.~(\ref{eq:n(r)}) was expressed as a trace
over the projected density operator:\cite{Baer2013}
\begin{equation}
n\left(\boldsymbol{r}\right)=2\text{Tr}\left[\theta\left(\mu-\hat{h}_{\text{KS}}\right)\delta\left(\boldsymbol{r}-\hat{\boldsymbol{r}}\right)\right].\label{eq:trace}
\end{equation}
The problem now shifts into calculating self-consistently the trace
in Eq.~(\ref{eq:trace}) (since $\hat{h}_{\text{KS}}$ depends on
$n\left(\boldsymbol{r}\right)$) rather than solving the KS equations
by brute-force diagonalization. When the trace is performed using
the KS eigenstates, the computational cost remains $O\left(N_{e}^{3}\right)$
similar to the traditional approach. However, since the trace is invariant
to the basis, alternative schemes that potentially lead to improved
scaling can be used. One such scheme is based on the concept of a
stochastic trace formula, which reduces the scaling of the trace operation
by introducing a controlled statistical error.\cite{Hutchinson1990} 

Using the stochastic trace formula, the density can be estimated as
a symmetrized stochastic trace formula, given by:\cite{Baer2013}

\begin{align}
n^{I}\left(\boldsymbol{r}\right) & =\left\langle \left\langle \chi\right|\sqrt{\theta_{\beta}\left(\mu-\hat{h}_{{\rm KS}}\right)}2\delta\left(\boldsymbol{r}-\hat{\boldsymbol{r}}\right)\right.\label{eq:stoch-dens}\\
 & \,\,\,\,\,\,\,\,\,\,\,\,\,\,\,\,\,\,\,\left.\times\sqrt{\theta_{\beta}\left(\mu-\hat{h}_{{\rm KS}}\right)}\left|\chi\right\rangle \right\rangle _{\chi}\nonumber 
\end{align}
where $\left\langle \cdots\right\rangle _{\chi}$ denotes an average
over $I$ stochastic orbitals $\left|\chi\right\rangle $, defined
as:
\begin{equation}
\left\langle \left.\mathbf{r}\right|\chi\right\rangle =h^{-3/2}e^{i\varphi_{\boldsymbol{r}}}
\end{equation}
for each grid point $\boldsymbol{r}$, the parameter $h$ (not to
be confused with the KS Hamiltonian $\hat{h}_{KS}$ operator) is the
grid spacing, and $\varphi_{\boldsymbol{r}}$ are statistically independent
random variables in the range $[0,2\pi]$ ($\left\langle e^{i\varphi_{\boldsymbol{r}}}e^{-i\varphi_{\boldsymbol{r}^{\prime}}}\right\rangle _{\varphi}=\delta_{\boldsymbol{r}\,\boldsymbol{r}^{\prime}}$).
The density $n\left(\boldsymbol{r}\right)$ is, strictly speaking,
given by the limit $n\left(\boldsymbol{r}\right)=\lim_{I\rightarrow\infty}n^{I}\left(\boldsymbol{r}\right)$
and we approximate it with a finite $I$. The Heaviside function in
Eq.~(\ref{eq:stoch-dens}) is smoothed by the function $\theta_{\beta}\left(\varepsilon\right)\equiv\frac{1}{2}\mbox{erfc}\left[\beta\varepsilon\right]$,
where $\beta$ is a large constant satisfying $\beta E_{g}\gg1$,
where $E_{g}$ is the KS-DFT fundamental gap. Throughout this paper
we set the value of $\beta$ to $100E_{h}^{-1}$. The action of $\sqrt{\theta_{\beta}\left(\mu-\hat{h}_{{\rm KS}}\right)}$
on $|\chi\rangle$ is evaluated by a Chebyshev expansion in powers
of the sparse KS Hamiltonian, $\hat{h}_{{\rm KS}}$.\cite{Tal-Ezer1984}
The length of the Chebyshev series is determined by the value of $\left(\mu-E_{min}\right)/\Delta E$
and $\beta\Delta E$ where $\Delta E=\left(E_{max}-E_{min}\right)/2$
and $E_{min/max}$ are the minimal and maximal eigenvalues of $\hat{h}_{KS}$.
Under the conditions of the present systems, the length of the series
is \textasciitilde{}3000 terms. 

The stochastic trace evaluation (Eq.~\ref{eq:stoch-dens}) reduces
the computational scaling of KS-DFT to $O\left(N_{e}\right)$ and
for certain properties even to a sub-linear scaling.\cite{Baer2013}
Linear-scaling complexity is achieved due to the following facts:
1) the application of a Hamiltonian to a stochastic orbital, $\hat{h}_{KS}\left|\chi\right\rangle $
requires a linear scaling effort (irrespective of the structure of
the orbital); 2) The length of the Chebyshev series is independent
(or at most weakly dependent) of system size and 3) Only a system-size
independent number of stochastic orbitals are required. This type
of assumptions is different from the linear-scaling approaches depending
on density matrix sparsity,\cite{Baer1997a,Goedecker1999} which assume
that locality of orbitals is not completely destroyed by the repeated
operation of the Hamiltonian.

A converged self-consistent solution of Eq.~(\ref{eq:stoch-dens})
provides an estimate of the electron density and in addition can be
used to generate other quantities, such as the density of states (DOS),
the total energy per electron, and the forces acting on the nuclei.
All estimates contain a statistical error that can be controlled by
increasing the number of stochastic orbitals ($I$) used to evaluate
the trace in Eq.~(\ref{eq:stoch-dens}). Of particular relevance
to this work are the Cartesian forces exerted by the electrons on
$N$ nuclei ($\alpha=1,\dots,N$), which can be evaluated through
the Hellmann-Feynman theorem:\cite{Hellman1937,Feynman1939} 
\begin{equation}
\boldsymbol{f}_{\alpha}=-\int\frac{\partial v_{eN}\left(\mathbf{r}\right)}{\partial\boldsymbol{R}_{\alpha}}n^{I}\left(\mathbf{r}\right)d^{3}r.\label{eq:hellman-Fey Force}
\end{equation}
It should be stressed that for finite sampling, these forces are only
approximately commensurate with the stochastic estimate of the energy
(which is not used in the sampling procedure at all), as discussed
in Appendix~\ref{sec:Hellmann-Feynman-Forces}. The stochastic estimate
of the Hellmann-Feynman forces is an excellent estimator of the deterministic
forces, as can be seen in Fig.~\ref{fig:std of frags}, where the
mean absolute deviation is dominated by the fluctuations and not by
additional bias terms.

These sDFT forces can be expressed as:
\begin{equation}
\boldsymbol{f}_{\alpha}=\boldsymbol{f}_{\alpha}^{\text{det}}+\boldsymbol{f}_{\alpha}^{\text{fluc}}+\boldsymbol{f}_{\alpha}^{\text{bias}}\label{eq:force-deviations}
\end{equation}
where $\boldsymbol{f}_{\alpha}^{\text{det}}$ is the deterministic
(generally unknown) force, $\boldsymbol{f}_{\alpha}^{\text{fluc}}$
is the pure fluctuating term, and $\boldsymbol{f}_{\alpha}^{\text{bias}}$
is the bias expected to be proportional to $\frac{1}{I}$ in leading
order. The choice of $I$ should be large enough to reduce $\boldsymbol{f}_{\alpha}^{\text{bias}}$
to negligible values and the only source of error in the procedure
is then the statistical fluctuations proportional to $\frac{1}{\sqrt{I}}$
with vanishing mean ($\left\langle \boldsymbol{f}_{\alpha}^{\text{fluc}}\right\rangle =0$).

\subsection{Embedded saturated fragments sDFT}

The reduction of the scaling in sDFT is achieved by replacing the
deterministic, numerically exact, trace evaluation with a stochastic
sampling of the density. In return, this leads to statistical errors
in the computed observables. To reduce the size of the statistical
fluctuations, an embedded saturated  fragments method is introduced
inspired by (but different from) the method of Ref.~\onlinecite{Neuhauser2014a}
. The latter approach was suitable mainly for systems composed of
proximate but chemically separated molecules (like clusters of water
molecules, for examples). The present method is applicable for fragmenting
covalently bonded systems, like silicon NCs. 

In this approach, the system is divided into $F$ small fragments
that are possibly overlapping. The division to fragments is flexible,
and any desired physically motivated fragmentation can be used. The
density is then a sum of the fragment density and a small correction
term:
\begin{equation}
n\left(\boldsymbol{r}\right)=n_{F}\left(\boldsymbol{r}\right)+\Delta n\left(\boldsymbol{r}\right)\label{eq:frag-corr-dens-normal-form}
\end{equation}
where $n_{F}\left(\boldsymbol{r}\right)=\sum_{f=1}^{F}n_{f}\left(\boldsymbol{r}\right)$
is the density generated by the individual fragments obtained from
a deterministic KS-DFT calculation for each fragment and $\Delta n\left(\boldsymbol{r}\right)=\left(n^{I}\left(\boldsymbol{r}\right)-n_{F}^{I}\left(\boldsymbol{r}\right)\right)$
is a correction term evaluated using stochastic orbitals. Here, $n^{I}\left(\boldsymbol{r}\right)$
is given by Eq.~(\ref{eq:stoch-dens}) and $n_{F}^{I}\left(\boldsymbol{r}\right)=\sum_{f=1}^{F}n_{f}^{I}\left(\boldsymbol{r}\right)$
is a sum over a \emph{stochastic }estimate of the fragments density.
In the limit $I\rightarrow\infty$, Eqs.~(\ref{eq:stoch-dens}) and
(\ref{eq:frag-corr-dens-normal-form}) are identical and equal to
the deterministic density. For finite values of $I$, the size of
the statistical fluctuations of the two approaches are quite different.
Since the deterministic fragmented density, $n_{F}\left(\boldsymbol{r}\right)$,
provides a reasonable approximation for the full density $n\left(\boldsymbol{r}\right)$,
the correction term, $\Delta n\left(\boldsymbol{r}\right)$, which
is evaluated stochastically, is rather small, leading to a reduced
variance in the relevant observables (forces, DOS, total energy per
electron, etc.) compared to the direct stochastic approach of Eq.~(\ref{eq:stoch-dens}).
An equivalent viewpoint is that the fragmentation is a device for
reducing the variance in the stochastic evaluation of the density.
This is evident by rewriting Eq.~(\ref{eq:frag-corr-dens-normal-form})
in the following form

\begin{equation}
n\left(\boldsymbol{r}\right)=n^{I}\left(\boldsymbol{r}\right)+\sum_{f=1}^{F}\left(n_{f}\left(\boldsymbol{r}\right)-n_{f}^{I}\left(\boldsymbol{r}\right)\right),\label{eq:frag-corr-dens-roi-form}
\end{equation}
and the implementation of this form is described in Appendix~(\ref{sec:The-embedded-fragments}).

To assess the accuracy of the embedded saturated  fragmented sDFT
(efsDFT), we calculated the standard deviations (STDs) and mean absolute
deviations with respect the deterministic DFT (MADs) of the atomic
forces in a $\text{Si}{}_{35}\text{H}_{36}$ NC using hydrogen passivated
$\text{Si}{}_{5}$ fragments. The results are shown in Fig.~\ref{fig:std of frags}.\footnote{All calculations in this work use real-space grids of spacing $\Delta x=0.5a_{0}$,
Troullier-Martins norm-conserving pseudopotentials~\cite{Troullier1991}
within the Kleinman-Bylander approximation.\cite{Kleinman1982} Fast
Fourier Transforms were used for applying the kinetic energy operator
and for determining the Hartree potentials and the method of Ref.~\onlinecite{Martyna1999}
was used for treating the long range Coulomb interactions in a finite
simulation cell with periodic boundary conditions. DFT calculations
were performed under the local density approximation (LDA). } The STDs and MADs decrease as $1/\sqrt{I}$, indicating that the
bias in the force estimation is negligible. The standard deviations
in the sDFT forces are larger by a factor of $\approx3$ compared
to those of efsDFT. This implies that the required number of stochastic
orbitals in efsDFT is nearly an order of magnitude smaller than in
sDFT for similar STDs. The STDs can be further reduced by using larger
fragments as discussed below (cf., Fig.~\ref{fig:std0-Frags}).

\subsection{\label{subsec:Langevin-dynamics-based}Langevin dynamics based on
efsDFT }

The standard approach to generate canonically distributed configurations
\emph{using ab initio} techniques is based on molecular dynamics,
which requires as input accurate force estimates for each atomic degree
of freedom. Since the forces generated by efsDFT contain a stochastic
component, we use Langevin dynamics (LD) instead of molecular dynamics
to sample configurations according to the Boltzmann distribution.
A LD trajectory~\cite{Allen1987,VanKampen1992,Zwanzig2001,Frenkel2002}
is a sequence of configurations $\left(\boldsymbol{p},\boldsymbol{q}\right)^{m}=\left(\boldsymbol{p}\left(t_{m}\right),\boldsymbol{q}\left(t_{m}\right)\right)$
at discrete ``times'' $t_{m}=m\Delta t$, where $\Delta t$ is the
time step, and $\boldsymbol{q}\equiv\left(\boldsymbol{q}_{1},\dots,\boldsymbol{q}_{N}\right)$
and $\boldsymbol{p}\equiv\left(\boldsymbol{p}_{1},\dots,\boldsymbol{p}_{N}\right)$
are the Cartesian coordinates and conjugate momenta, respectively,
for the $N$ atoms, The trajectory is a solution of the Langevin equation
(LE) of motion:\cite{Langevin1908}

\begin{equation}
\mu_{\alpha}\ddot{\boldsymbol{q}}_{\alpha}=\boldsymbol{f}_{\alpha}\left(\boldsymbol{q}\right)-\gamma_{\alpha}\boldsymbol{p}_{\alpha}+\boldsymbol{\eta}_{\alpha}.\label{eq:Langevin eq}
\end{equation}
where $\mu_{\alpha}$ is the mass of the atom $\alpha$, $\gamma_{\alpha}$
is its friction constant, and $\boldsymbol{f}_{\alpha}=\boldsymbol{f}_{\alpha}^{\text{det}}+\boldsymbol{f}_{\alpha}^{\text{fluc}}$
is the total efsDFT force acting on it, including deterministic and
fluctuating parts (see Eq.~\ref{eq:force-deviations}). The bias
is assumed negligible, so that $\left\langle \boldsymbol{f}_{\alpha}\right\rangle =\boldsymbol{f}_{\alpha}^{\text{det}}$
. In Eq.~(\ref{eq:Langevin eq}) $\boldsymbol{\eta}_{\alpha}$ is
an additional uncorrelated white-noise force introduced so as to satisfy
the fluctuation-dissipation (FD) relation. We require that the total
random fluctuation on each atom obey:
\begin{align*}
\left\langle \boldsymbol{\eta}{}_{\alpha}\left(t\right)\right\rangle  & =\left\langle \boldsymbol{f}_{\alpha}^{\text{fluc}}\left(t\right)\right\rangle =0
\end{align*}
and
\begin{align}
\left\langle \left(\boldsymbol{\eta}{}_{\alpha}\left(t\right)+\boldsymbol{f}_{\alpha}^{\text{fluc}}\left(t\right)\right)\otimes\left(\boldsymbol{\eta}{}_{\alpha'}\left(t^{\prime}\right)+\boldsymbol{f}_{\alpha'}^{\text{fluc}}\left(t^{\prime}\right)\right)\right\rangle  & =\nonumber \\
\left\langle \boldsymbol{\eta}{}_{\alpha}\left(t\right)\otimes\boldsymbol{\eta}{}_{\alpha'}\left(t^{\prime}\right)\right\rangle +\left\langle \boldsymbol{f}_{\alpha}^{\text{fluc}}\left(t\right)\otimes\boldsymbol{f}_{\alpha'}^{\text{fluc}}\left(t^{\prime}\right)\right\rangle  & =\label{eq:fluc-correlation}\\
\text{\ensuremath{\boldsymbol{\text{I}}_{3\times3}}}\sigma_{\alpha}^{2}\delta_{\alpha\alpha'}\delta\left(t-t^{\prime}\right),\nonumber 
\end{align}
where $\alpha,\alpha^{\prime}=1,\dots N$ are atom indices, $\left\langle \cdots\right\rangle $
designates average over the atomic force distribution, $\boldsymbol{\text{I}}_{3\times3}$
is the $3\times3$ unit matrix, and $\sigma_{\alpha}$ is the atomic
force STD of atom $\alpha$, which is taken to satisfy the fluctuation-dissipation
relation:

\begin{equation}
\sigma_{\alpha}^{2}=2\mu_{\alpha}\gamma_{\alpha}k_{{\rm B}}T.\label{eq:FDT}
\end{equation}

We use the Verlet-like algorithm~\cite{Gronbech-Jensen2013} for
numerically integrating the LE of motion at a fixed temperature $T$
and a predefined time-step $\Delta t$. The positions and momenta
in time step $m+1$ depend on the positions and momenta in time step
$m$ as well as on the forces in time step $m$ and the additional
white noise $\boldsymbol{\eta}_{\alpha}^{m}$ is sampled from a Gaussian
distribution such that the discretized version of Eq.~(\ref{eq:fluc-correlation})
holds: $\left\langle \left(\boldsymbol{\eta}{}_{\alpha}^{m}+\boldsymbol{f}_{\alpha}^{\text{fluc}}\right)\otimes\left(\boldsymbol{\eta}{}_{\alpha'}^{n}+\boldsymbol{f}_{\alpha'}^{\text{fluc}}\right)\right\rangle \Delta t=\text{\ensuremath{\boldsymbol{\text{I}}}}\sigma_{\alpha}^{2}\delta_{\alpha\alpha'}\delta_{mn}$:
\begin{align}
\boldsymbol{q}_{\alpha}^{m+1} & =\boldsymbol{q}_{\alpha}^{m}+b_{\alpha}\Delta t\mu_{\alpha}^{-1}\boldsymbol{p}_{\alpha}^{m}+\frac{1}{2}b_{\alpha}\Delta t^{2}\mu_{\alpha}^{-1}\left(\boldsymbol{f}_{\alpha}^{m}+\boldsymbol{\eta}_{\alpha}^{m+1}\right)\nonumber \\
\boldsymbol{p}_{\alpha}^{m+1} & =a_{\alpha}\boldsymbol{p}_{\alpha}^{m}+\frac{1}{2}\Delta t\left(a_{\alpha}\boldsymbol{f}_{\alpha}^{m}+\boldsymbol{f}_{\alpha}^{m+1}+2b_{\alpha}\boldsymbol{\eta}_{\alpha}^{m+1}\right),\label{eq:Discretized-LE}
\end{align}
where $a_{\alpha}=b_{\alpha}\left(1-\frac{1}{2}\gamma_{\alpha}\Delta t\right)$
and $b_{\alpha}^{-1}=1+\frac{1}{2}\gamma_{\alpha}\Delta t$. The algorithm
allows for stable and accurate solutions of the LE with time step
comparable to that used in molecular dynamics simulations for similar
systems. It treats the additional white noise component $\boldsymbol{\eta}_{\alpha}$
of the force differently from the force $\boldsymbol{f}_{\alpha}=\boldsymbol{f}_{\alpha}^{\text{det}}+\boldsymbol{f}_{\alpha}^{\text{fluc}}$
that result from the efsDFT calculation containing deterministic and
fluctuating components that cannot be separated.

\begin{figure}
\begin{centering}
\includegraphics[width=0.9\columnwidth]{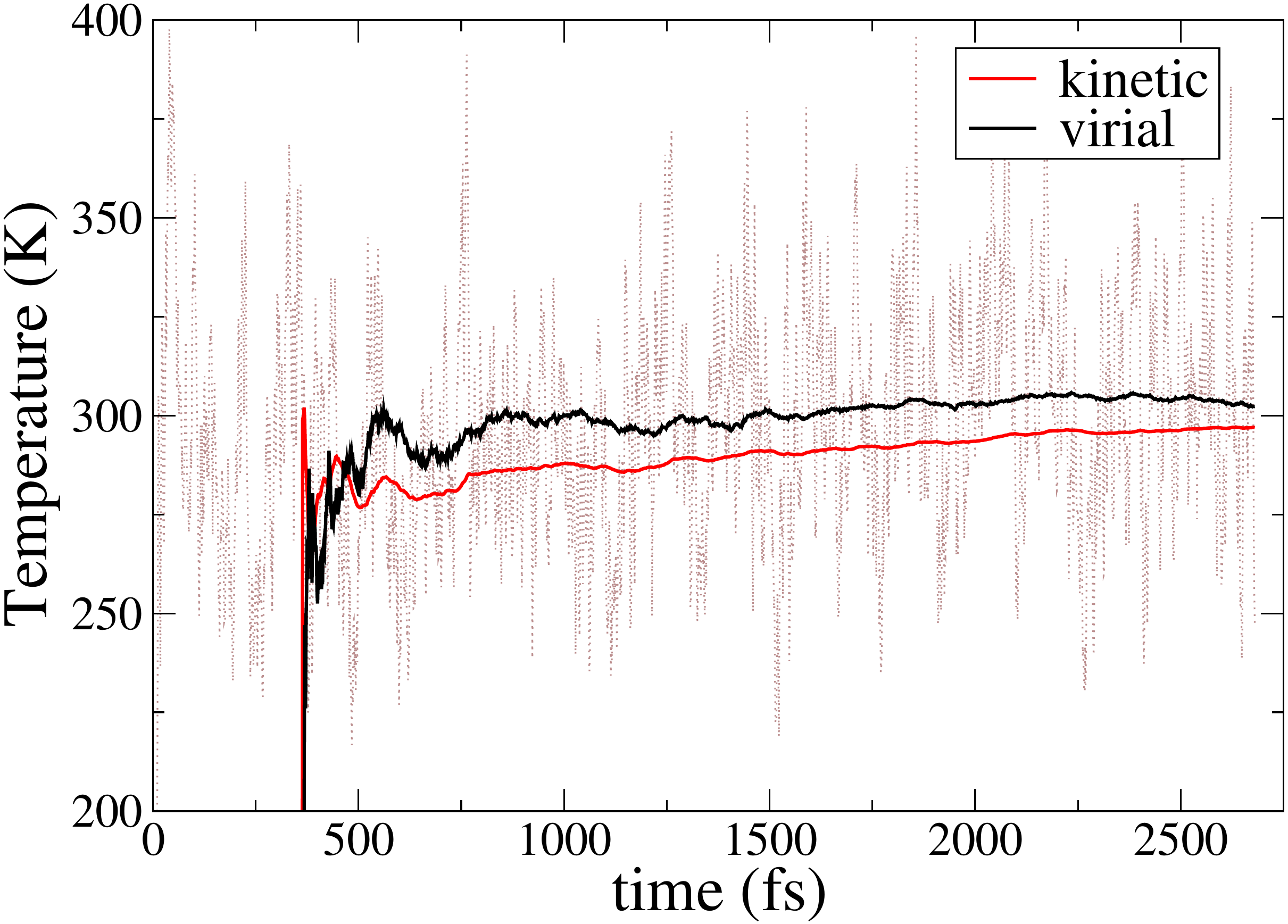}
\par\end{centering}
\caption{\label{fig:temperature}The Verlet-like \cite{Gronbech-Jensen2013}
temperature in $\text{Si}_{35}\text{H}_{36}$ evaluated with efsDFT
using: transient kinetic energy (dotted line), running average kinetic
energy for Si (red) and the running average virial (black) using $I=30$
stochastic orbitals, a time step of $\Delta t=1.2\mbox{\,fs}$ and
the friction coefficients $\gamma_{\text{Si}}=\gamma_{\text{H}}=0.04\,\mbox{fs}^{-1}$.
In the canonical distribution, the average kinetic energy $T$ is
equal to $\sqrt{\frac{3N}{2}}\delta T$ where $\delta T$ is the fluctuation
in the kinetic energy.\cite{Martinez2015} The standard deviation
of the transient shown as a dotted line is $\delta T=29$K and multiplied
by $\sqrt{3N/2}$, where $N=71$ is the number of atoms in the system,
gives $299.3K$, which is close to the designated temperature. The
average kinetic energy of Si and H are 315K and 285K respectively. }
\end{figure}

In Fig.~\ref{fig:temperature} we plot for $\text{Si}_{35}\text{H}_{36}$
the running average of the transient temperature, $T^{m}$, calculated
from the kinetic energy
\begin{equation}
T_{K}^{m}=\frac{2}{3Nk_{{\rm B}}}\sum_{\alpha}\left(\boldsymbol{p}_{\alpha}^{m}\right)^{2}/2\mu_{\alpha}
\end{equation}
 and from the virial estimator, 
\begin{equation}
T_{V}^{m}=-\frac{1}{3Nk_{{\rm B}}}\sum_{\alpha}\left(\boldsymbol{f}_{\alpha}^{m}+\boldsymbol{\eta}_{\alpha}^{m}\right)\cdot\left(\boldsymbol{q}_{\alpha}^{m}-\left\langle \boldsymbol{q}_{\alpha}\right\rangle \right).
\end{equation}
In the above, $\left\langle \boldsymbol{q}_{\alpha}\right\rangle $
is the time average of the coordinate of atom $\alpha$. The initial
positions of the Si atoms were taken from the bulk values. All surface
Si atoms with more than two dangling bonds were removed and the remaining
surface Si atoms were passivated using one or two H atoms placed in
a tetrahedral geometry at the Si-H distance of $1.47$Å. The momenta
were sampled from a Boltzmann distribution at $T=300K$. This non-equilibrium
initial configuration relaxes towards equilibrium. 

The agreement in Fig.~\ref{fig:temperature} between the two temperature
estimators is consistent with a proper sampling of the canonical distribution
of both positions and velocities. The small discrepancies at the longest
averaging time are due to the large fluctuations of the transient
temperature, particularly when using the virial estimator. We have
also calculated the fluctuations in the kinetic energy and found good
agreement with the corresponding analytical value (see caption of
the figure). The two atomic species have a slightly $\pm5\%$ deviations
in the temperatures. These may result from several factors such as
the finite timestep of the Langevin propagator \cite{Gronbech-Jensen2013},
incomplete SCF convergence,\cite{Martinez2015} insufficiently accurate
estimate of the amount of white noise $\eta_{\alpha}^{m}$ in Eq.~(\ref{eq:fluc-correlation})
required to fulfill the fluctuation-dissipation relation.

\subsection{\label{subsec:friction}Determining the optimal friction }

\begin{figure}
\includegraphics[width=0.7\columnwidth]{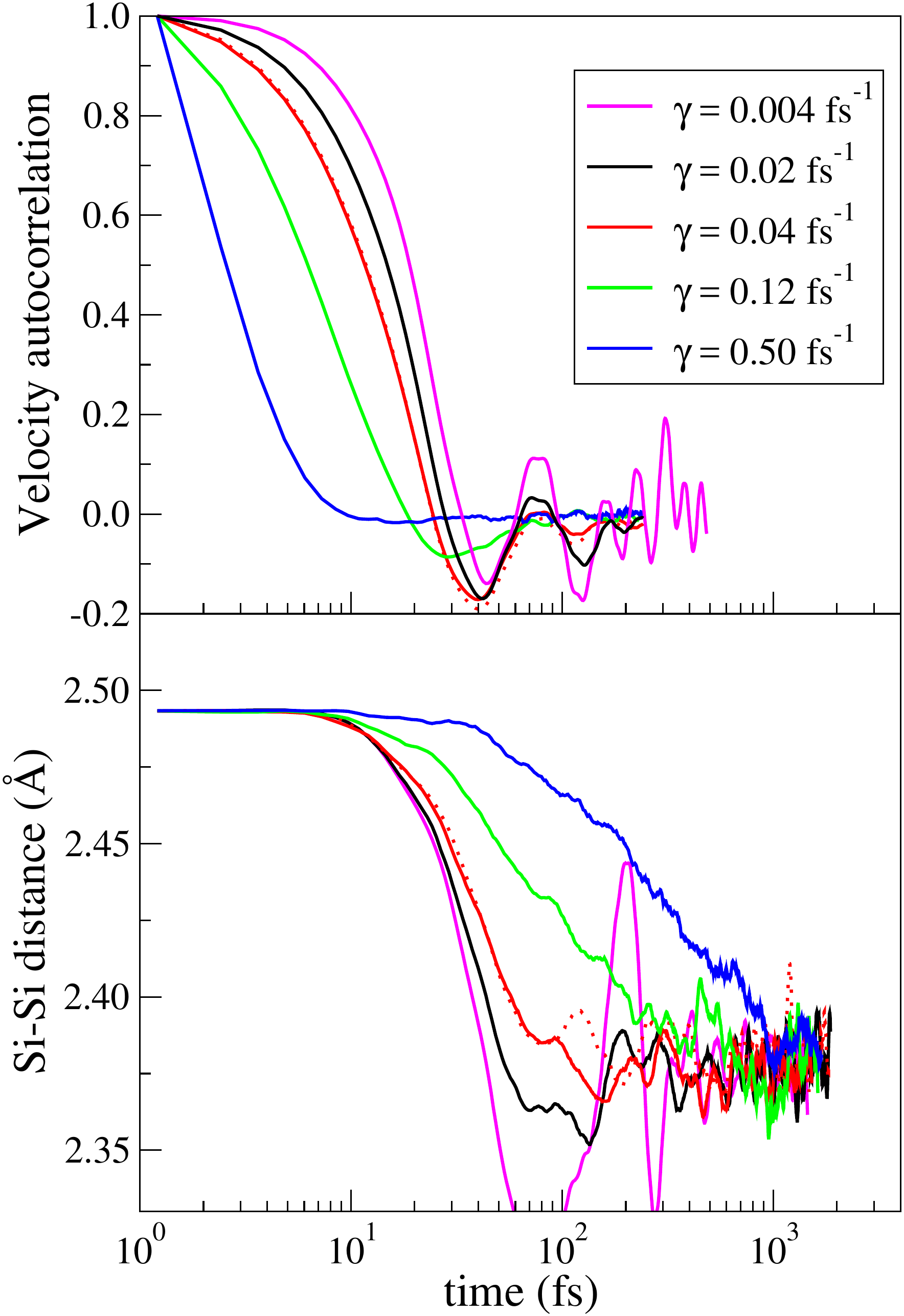}\caption{\label{fig:thermalization-times}The normalized velocity autocorrelation
function (top panel) and the mean nearest-neighbor Si-Si distance
(bottom panel) in $\text{Si}_{35}\text{H}_{36}$ as a function of
time for a LD trajectory at $T=300K$ with time step $\Delta t=1.2\text{\,fs}$
calculated using a dDFT based LD for different values of $\gamma=\gamma_{\text{Si}}=\gamma_{\text{H}}$.
The dashed curve corresponds to a efsDFT based LD calculation with
$\gamma=0.04\,\text{fs}^{-1}$ and $I=30$ stochastic orbitals. The
simulation started intentionally from an inflated configuration in
order to to measure the relaxation time.}
\end{figure}

\begin{figure}[t]
\begin{centering}
\includegraphics[width=0.9\columnwidth]{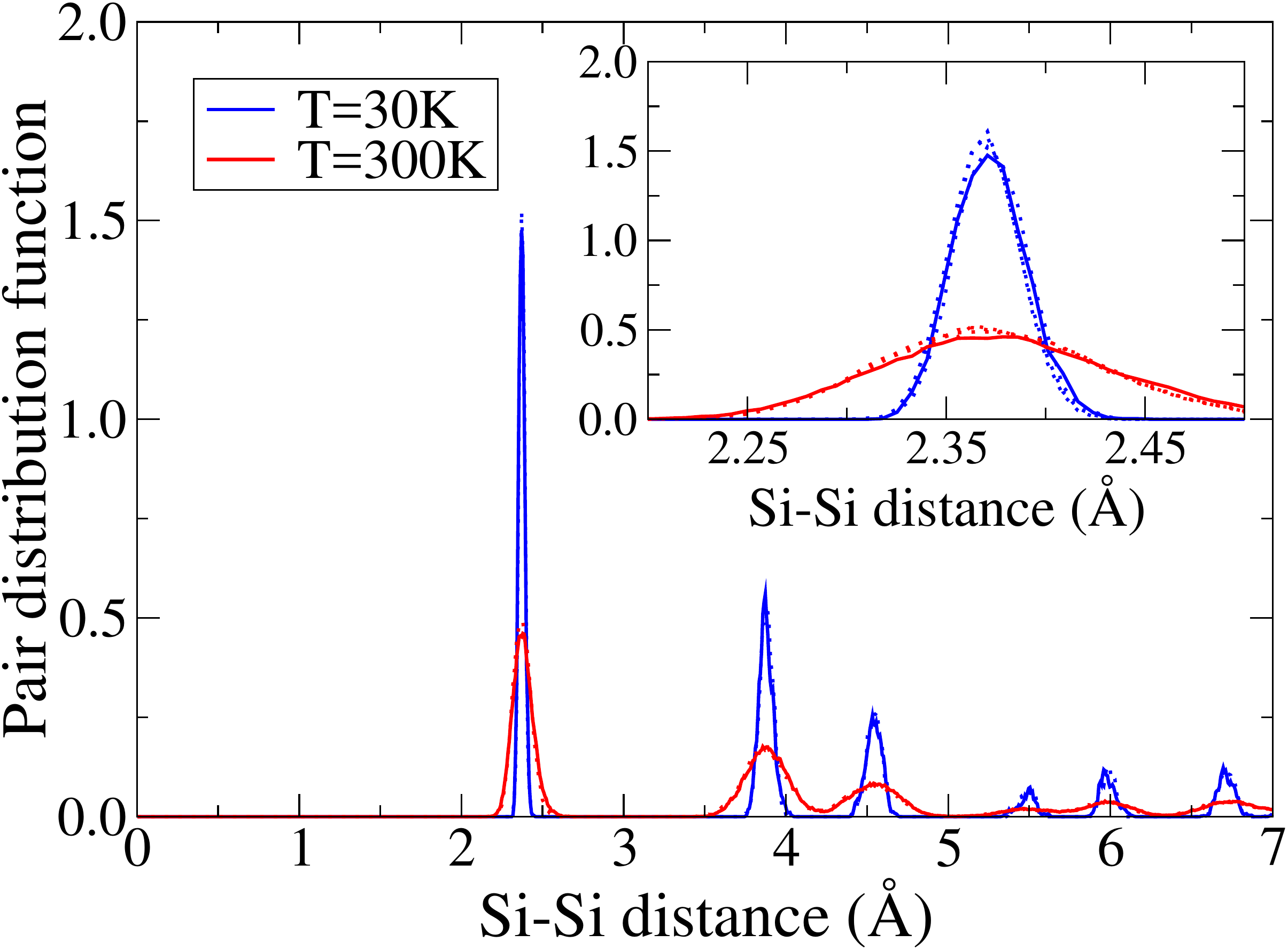}
\par\end{centering}
\caption{\label{fig:g(r) Si35}The Si-Si pair distribution function $g\left(r\right)$
for $\text{S\ensuremath{i_{35}H_{36}}}$ calculated using dDFT (dotted
lines) and efsDFT based LD (solid lines, see Table~\ref{tab:Params}
for parameters) at $T=30\text{K}$ , (blue curves) and $300\text{K}$
(red curves). Inset: Details of the first (nearest neighbor) peak.
The dotted lines are for different friction coefficients $\gamma=\gamma_{\text{Si}}=\gamma_{\text{H}}$
in the range $0.02-0.5\,\text{fs}^{-1}$ . }
\end{figure}

The effect of $\gamma_{\text{Si}}$ on the configurational relaxation
and on the velocity autocorrelation decay is illustrated in Fig.~\ref{fig:thermalization-times}
for $\text{Si}_{36}\text{H}_{35}$. In order to decrease the number
of unknown parameters we set the values of $\gamma_{\alpha}$ and
$\sigma_{\alpha}$ to be identical for all atoms of the same type
(i.e. Si or H in the systems studied here). To achieve such a uniform
value of $\sigma$ we introduced white noise $\eta_{\alpha}^{i}=\sqrt{\left(\sigma_{\alpha}^{i}\right)^{2}-\left\langle \left(f_{\alpha}^{i,\text{fluc}}\right)^{2}\right\rangle }$
for each degree of freedom (see Eq.~\ref{eq:fluc-correlation}),
where $\left\langle \left(f_{\alpha}^{i,\text{fluc}}\right)^{2}\right\rangle $
is estimated by a separate set of runs on the initial NC configuration
using several independent sets of stochastic orbitals. Note, that
we have tested that the magnitude of the sDFT force fluctuation $\left\langle \left(f^{\text{fluc}}\right)^{2}\right\rangle $
is not sensitive to the particular configuration used.

\begin{table}[H]
\begin{centering}
\begin{tabular}{|c|c|c|c|c|c|c|}
\hline 
\multirow{2}{*}{NC} & \multirow{2}{*}{T(K)} & \multicolumn{2}{c|}{$\gamma\left(\text{fs}^{-1}\right)$} & \multirow{2}{*}{$I$} & \multirow{2}{*}{$\Delta t\left(\text{fs}^{-1}\right)$} & \multirow{2}{*}{$t_{\text{iter}}$$\left(\text{min}\right)$}\tabularnewline
\cline{3-4} 
 &  & H & Si &  &  & \tabularnewline
\hline 
\multirow{2}{*}{$\text{Si}_{35}\text{H}_{36}$} & \multicolumn{1}{c|}{30} & 0.12 & 0.04 & 120 & 1.2 & 1\tabularnewline
\cline{2-7} 
 & \multicolumn{1}{c|}{300} & 0.04 & 0.04 & 30 & 1.2 & 1\tabularnewline
\hline 
\multirow{2}{*}{$\text{Si}_{147}\text{H}_{100}$} & 30 & 0.12 & 0.04 & 120 & 1.2 & 2\tabularnewline
\cline{2-7} 
 & 300 & 0.12 & 0.04 & 30 & 1.2 & 2\tabularnewline
\hline 
\multirow{2}{*}{$\text{Si}_{705}\text{H}_{300}$} & \multicolumn{1}{c|}{30} & 0.12 & 0.04 & 120 & 1.2 & 10\tabularnewline
\cline{2-7} 
 & \multicolumn{1}{c|}{300} & 0.12 & 0.04 & 92 & 1.2 & 10\tabularnewline
\hline 
\end{tabular}
\par\end{centering}
\caption{\label{tab:Params}Value of various parameters for the LD based on
efsDFT calculations: The friction coefficients $\gamma$, number of
stochastic orbitals $I$, time-step $\Delta t$ and the wall time
per single SCF iteration $t_{\text{iter}}$. }
\end{table}

As expected, the configurational relaxation time increases with increasing
values of $\gamma_{\text{Si }}$ with the opposite trend for the decay
time of the velocity autocorrelation function. Based on the results
of $\text{Si}_{36}\text{H}_{35}$ presented in Fig.~\ref{fig:thermalization-times}
we conclude that a friction coefficient of $0.04\text{fs}^{-1}$ is
sufficiently small for this system, with respect to minimizing both
velocity and pair distance autocorrelation times. Although lower values
of the friction coefficients could decrease the correlation time further,
they would require reducing the statistical noise, which would be
expensive to achieve using sDFT. Thus we chose the friction constants,
$\gamma_{Si}=0.04\text{fs}^{-1}$ for Silicon and $\gamma_{H}=0.12\text{fs}^{-1}$
for the lighter H atoms. These values were used for the larger systems
described in the next section (see Table~\ref{tab:Params}). Note
that the results shown in Fig.~\ref{fig:thermalization-times}, which
were generated using LD under dDFT, could have been equally well generated
under efsDFT. This is shown explicitly for $\gamma_{\text{Si}}=0.04\,\text{fs}^{-1}$
(dotted red line) proving that the relaxation times are similar to
those of the dDFT based LD calculation with the same value of $\gamma_{\text{Si}}$.

\subsection{\label{subsec:validation}Validation of LD within efsDFT }

Validation of the structure obtained using efsDFT based LD is demonstrated
using the pair distribution function $g\left(r\right)$.\cite{Allen1987}
For finite size NCs the average number of neighbors at a distance
$r$ is expected to be smaller than the bulk value due to surface
atoms with a smaller number of neighbors. Fig.~\ref{fig:g(r) Si35}
shows a close agreement between the dDFT and efsDFT based LD estimates
of $g\left(r\right)$ of the $\text{Si}_{35}\text{H}_{36}$ NC at
two temperatures. The inset focuses on the first peak in $g\left(r\right)$,
comparing the efsDFT to dDFT at $T=30$ and $300\text{K}$ .

\section{\label{sec:Results}Results}

\begin{figure}[t]
\includegraphics[width=0.9\columnwidth]{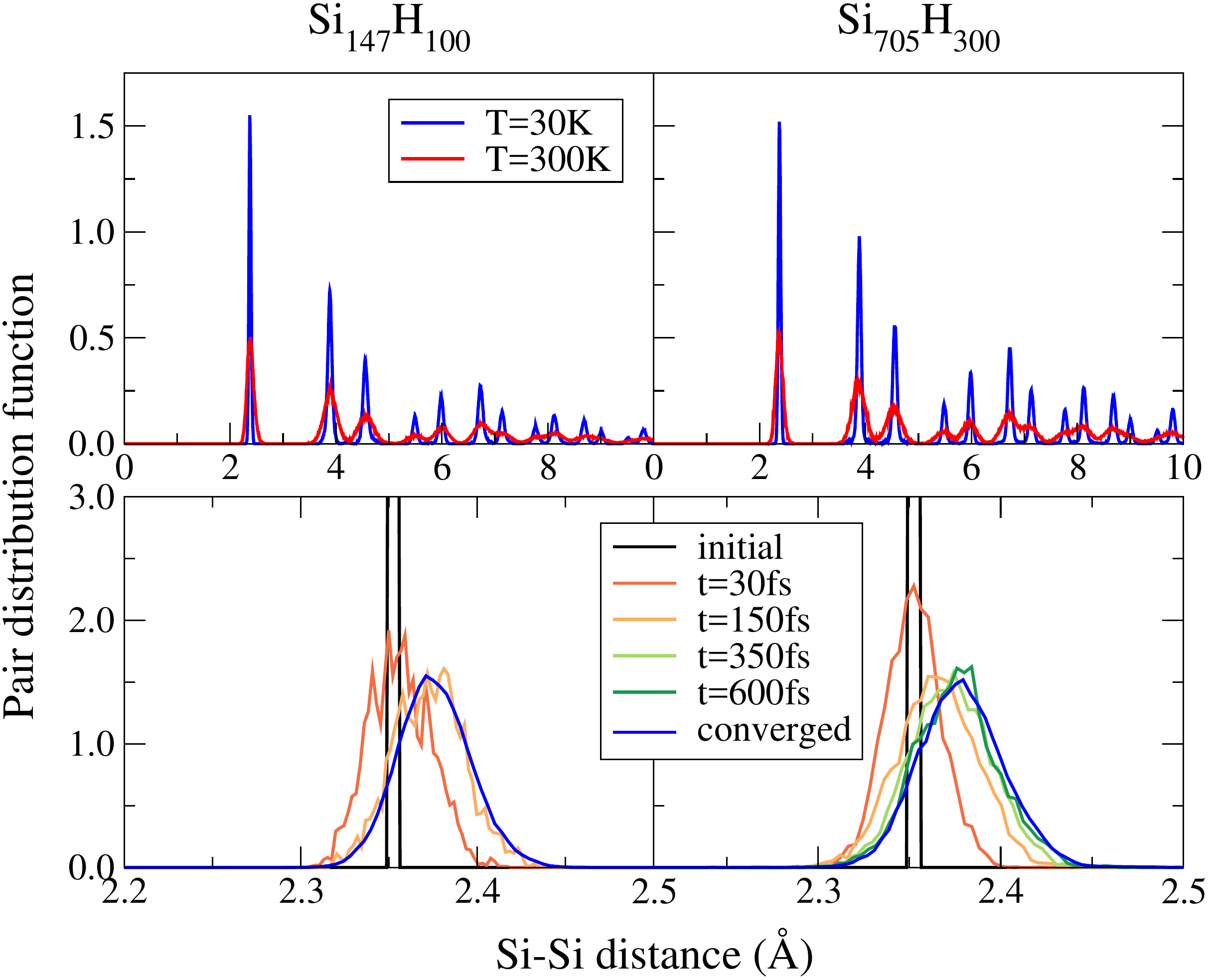}\caption{\label{fig:g(r) Si147 and Si705}The Si-Si pair distribution function
$g\left(r\right)$ for $\text{Si}_{147}\text{H}_{100}$ (left) and
$\text{Si}{}_{705}\text{H}_{300}$ (right) calculated using efsDFT
based LD. Upper panels: $g\left(r\right)$ for $T=30\text{K}$ (blue
curves) and $T=300\text{K}$ (red curves). Lower panels: The first
peak of $g\left(r\right)$ at $30\text{K}$ shown for several times.
The calculation parameters given in Table~\ref{tab:Params}. }
\end{figure}

In the previous sections we presented the methods and assessed the
accuracy and validity of the efsDFT based LD. Here we apply the method
to study structural properties of larger NCs exceeding $N_{e}=3000$
electrons. The Si-Si pair distribution functions $g\left(r\right)$
at two temperatures $T$ ($30$ and $300\text{K}$) are displayed
in the upper panel of Fig.~\ref{fig:g(r) Si147 and Si705} for $\text{Si}_{147}\text{H}_{100}$
and $\text{Si}_{705}\text{H}_{300}$. Temperature broadens the peaks
by a factor of $2-3$ without significantly changing the peak position. 

In the lower panel of Fig.~\ref{fig:g(r) Si147 and Si705} we plot
the transient and relaxed $g\left(r\right)$ at $30\text{K}$ for
the two systems, focusing on the first, nearest neighbor peak. As
described also for $\text{Si}_{35}\text{H}_{36}$, the initial positions
of the Si atoms for both systems were taken from the experimental
bulk values and all surface Si atoms with more than two dangling bonds
were removed. The remaining surface Si atoms were then passivated
using one or two H atoms placed in a tetrahedral position at the Si-H
distance of $1.47$Å. The initially sharp peak broadens and shifts
to longer Si-Si bond lengths as the system relaxes towards thermal
equilibrium. For $30\text{K}$, the relaxation times are $180$ and
$650\,$fs for $\text{Si}_{147}\text{H}_{100}$ and $\text{Si}_{705}\text{H}_{300}$,
respectively. For $300\text{K}$ they are $180$ and $250\,$fs respectively. 

\begin{figure}[t]
\includegraphics[width=1\columnwidth]{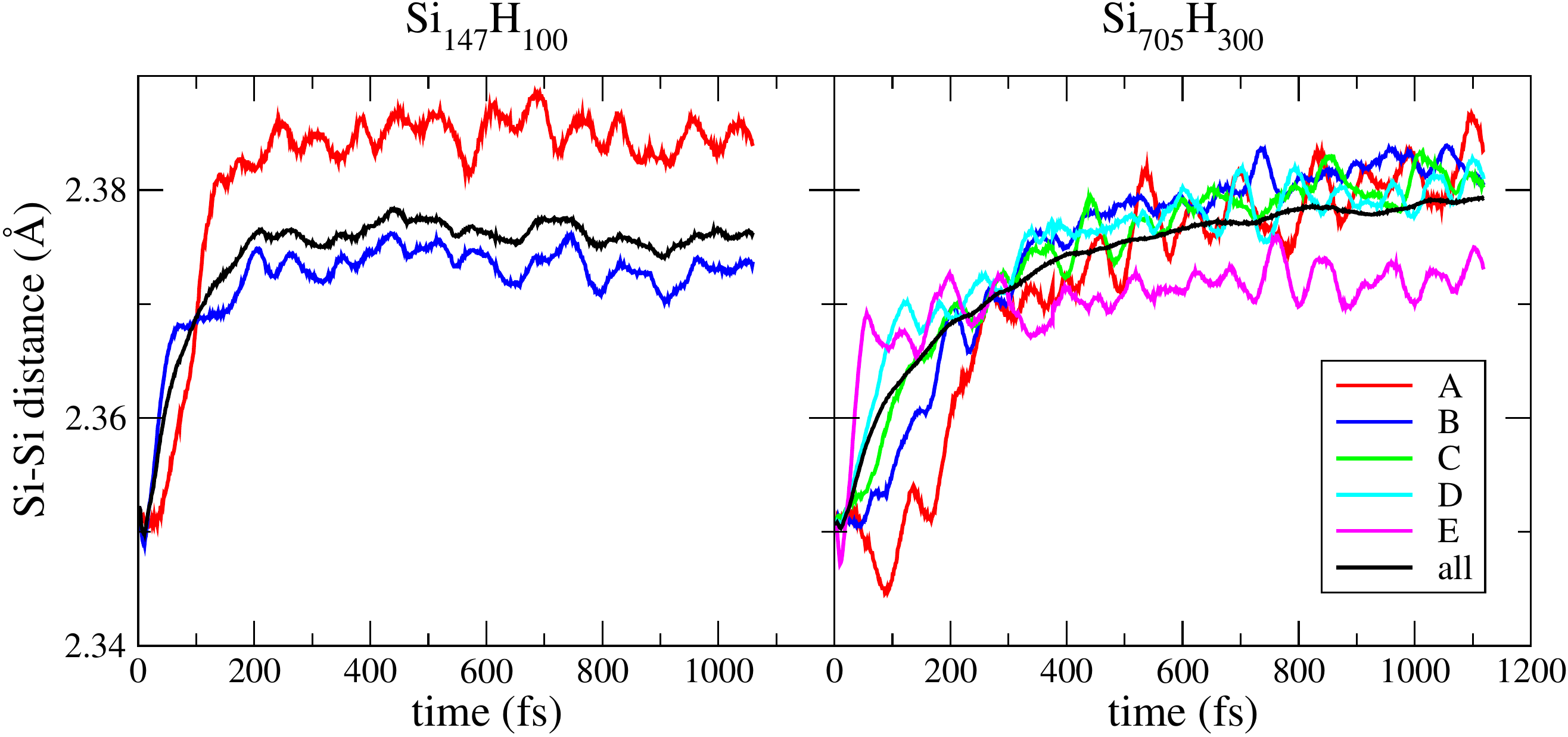}

\caption{\label{fig:Si-Si-Dist} Si-Si nearest neighbor distance averaged over
atoms in shells A-E (see Table~\ref{tab:Shells} for definition),
at $30\text{K}$ for $\text{Si}_{147}\text{H}_{100}$ (left) and $\text{Si}_{705}\text{H}_{300}$
(right) as a function of time. The calculation parameters given in
Table~\ref{tab:Params}. }
\end{figure}

\begin{table}[H]
\begin{centering}
\begin{tabular}{|c|c|c|c|c|}
\hline 
Shell & $R_{\text{in}}$ & $R_{\text{out}}$ & $N_{\text{Si}}$ & $N_{\text{NN}}$\tabularnewline
\hline 
\hline 
$A$ & 0 & 5.5 & 35 & 52\tabularnewline
\hline 
$B$ & 5.5 & 8.5 & 113 & 158\tabularnewline
\hline 
$C$ & 9.0 & 11.6 & 153 & 163\tabularnewline
\hline 
$D$ & 11.6 & 13.6 & 200 & 189\tabularnewline
\hline 
$E$ & 13.6 & 15.1 & 205 & 168\tabularnewline
\hline 
\end{tabular}
\par\end{centering}
\caption{\label{tab:Shells}The shells of the silicon NCs used for analyzing
the bond length relaxation in Fig.~\ref{fig:Si-Si-Dist}: Their inner
and outer radii (in Å), the number of Si atoms $N_{\text{Si}}$, and
the number of nearest neighbor (NN) Si-Si pairs $N_{\text{NN}}$.}
\end{table}

The relaxation transient is studied in greater detail in Fig.\ref{fig:Si-Si-Dist},
where the average nearest-neighbor bond lengths are shown for $\text{Si}_{147}\text{H}_{100}$
(spherical shells A-B) and $\text{Si}_{705}\text{H}_{300}$ (shells
A-E); see Table~\ref{tab:Shells} for the definition and properties
of the shells. In $\text{Si}_{705}\text{H}_{300}$ the deep layer
shells (A-D) relax slower than those near the surface showing that
relaxation progresses from the surface inwards. The difference between
the relaxation times of the two systems is correlated with the smaller
frequency, $\omega$, of the breathing mode of the larger NC. In the
limit of an over damped motion (as is the case here since $\gamma^{2}\gg\omega^{2}$),
the relaxation is dominated by two timescales proportional to $\gamma^{-1}$
and $\left(\omega^{2}/\gamma\right)^{-1}$. The former leads to a
fast relaxation while the latter is slower and depends on the value
of $\omega^{-2}$. The ratio of the breathing mode frequency for the
two particles is $\frac{\omega_{\text{L}}^{2}}{\omega_{\text{S}}^{2}}\approx2.8$
(L/S for large/small) assuming that the breathing mode frequency scales
linearly with the NC diameter.\cite{ghavanloo2015radial} This is
similar to the ratio of the relaxation times ($650/180=3.6$) for
the lower temperature. At the higher temperature, one needs to consider
anharmonic effects which are more pronounced in the large NC with
lower acoustic phonons.  Another noticeable feature in Fig.~\ref{fig:Si-Si-Dist}
is that the Si-Si bonds seem slightly shorter in $\text{Si}_{147}\text{H}_{100}$
than in $\text{S\ensuremath{i_{705}}}\text{\ensuremath{H_{300}}}$.
This results from the difference in the bond distance of atoms in
the outer shell, while the inner shell atoms have similar bond distances.

\section{\label{sec:Conclusions}Conclusions}

In this paper we developed an \emph{ab initio} Langevin dynamics approach
based on a new embedded saturated fragment stochastic DFT method.
We showed how the noisy forces resulting from the efsDFT calculation
are used to generate a set of configurations that are distributed
canonically at cryogenic and room temperatures. By proper choice of
the friction coefficients and the number of stochastic orbitals, thermalization
is reached within $\approx100$ time steps for these materials, since
the method is trivially parallelizable, larger computer resources
we allow to easily reduce the friction coefficients thus greatly improving
the sampling efficiency. While the methods presented here have already
allowed impressive achievements, such as determining structural properties
of silicon NCs of $3\text{nm}$ diameter containing more than $3000$
electrons, larger systems still, of unprecedented size, are now coming
within our grasp due to the fact that linear-scaling highly parallelizable
features of sDFT.
\begin{acknowledgments}
E.R. acknowledges support from the Physical Chemistry of Inorganic
Nanostructures Program, KC3103, Office of Basic Energy Sciences of
the United States Department of Energy under Contract DE-AC02-05CH11232.
D.N. acknowledge support by the NSF Grant DMR/BSF-1611382. R.B. acknowledges
the US-Israel Binational Science foundation support under the BSF-NSF
program, Grant 2015687.
\end{acknowledgments}

\appendix

\section{\label{sec:The-embedded-fragments}The embedded saturated fragments
approach}

\begin{figure*}[t]
\includegraphics[width=0.8\textwidth]{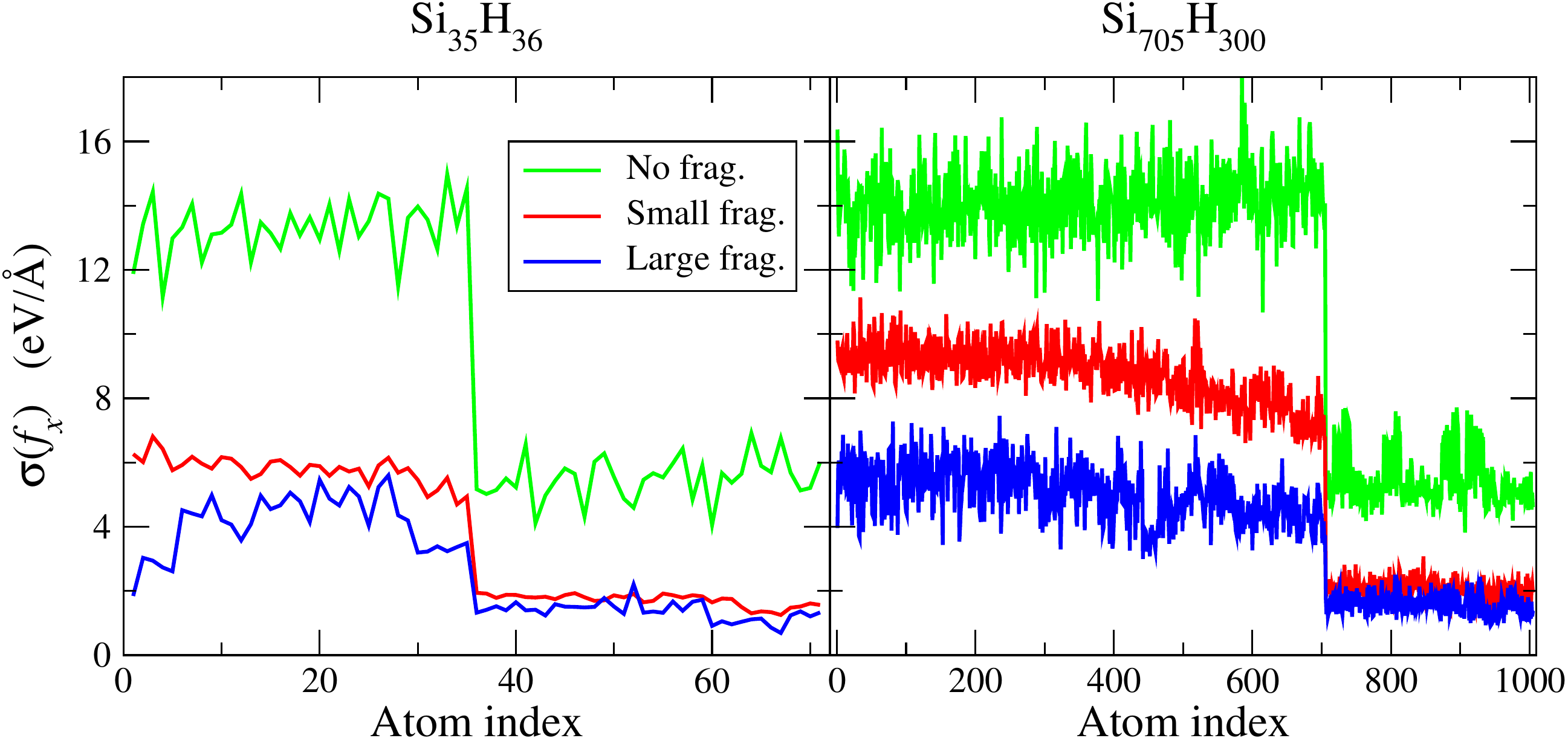}\caption{\label{fig:std0-Frags}Efficacy of fragments on the inherent sDFT
STD $\sigma_{1}\left(f_{x}\right)$ of the $x$-component of the force
on each atom of $\text{Si}_{35}\text{H}_{36}$ (left) and $\text{Si}_{705}\text{H}_{300}$
(right) NCs. The inherent STD $\sigma_{1}$ is the actual STD $\sigma$
times $\sqrt{I}$. Si atoms are shown first followed by H atoms, where
atoms are ordered by distance from the NC center. Calculations are
done on the configuration cut out from the bulk silicon, where H atoms
were placed near the surface for saturating the dangling bonds.}
\end{figure*}

Here, we provide the technical details for the embedded fragment method
described in Section~\ref{sec:Methods}. The method corrects the
stochastic estimate $\left\langle \hat{A}\right\rangle ^{I}$ for
the expectation value $\left\langle \hat{A}\right\rangle $ of a one-body
operator $\hat{A}$, using calculations performed on $F$ separate
fragments (see Eq.~(\ref{eq:frag-corr-dens-roi-form})):
\begin{equation}
\left\langle \hat{A}\right\rangle =\left\langle \hat{A}\right\rangle ^{I}+\sum_{f=1}^{F}\Delta A_{f}^{I},
\end{equation}
where the stochastic correction due to fragment $f$ is 
\begin{equation}
\Delta A_{f}^{I}=\left\langle \hat{A}_{f}\right\rangle -\left\langle \hat{A}_{f}\right\rangle ^{I},\label{eq:fragment-correction-A}
\end{equation}
 and the deterministic and the stochastic estimates, $\left\langle \hat{A}_{f}\right\rangle $
and $\left\langle \hat{A}_{f}\right\rangle ^{I}$ are calculated directly
on the fragment itself. 

Previous implementations of embedded fragment sDFT were applied to
systems of many weakly interacting molecules where the selection of
fragments or clusters of such molecules was natural.\cite{Neuhauser2014a}
We now describe a new method for defining and carrying calculations
with fragments which can break up covalently bonded systems such as
silicon NCs. The large system is divided into $F$ small fragments
composed of one or more bonded atoms each. The surface dangling bonds
of the fragment are passivated using a H atom placed in $1.46$ Å
from the Si atom, in the direction of the neighboring atom which is
not included in the fragment. This forms a saturated fragment. For
a saturated fragment $f$, the deterministic KS-DFT method is applied
to determine the KS eigenvalues $\varepsilon_{n}^{f}$ and eigenfunctions
$\psi_{n}^{f}\left(\boldsymbol{r}\right)$. Further, occupation numbers
$\left(p_{n}^{f}\right)^{2}=\frac{1}{2}\text{erfc}\left(\beta\left(\varepsilon_{n}^{f}-\mu_{f}\right)\right)$
are introduced for determining the saturated fragment density $n_{sf}\left(\boldsymbol{r}\right)=\sum_{n}\left(p_{n}^{f}\right)^{2}\psi_{n}^{f}\left(\boldsymbol{r}\right)^{2}$.
The fragment density $n_{f}\left(\boldsymbol{r}\right)=c_{f}\left(\boldsymbol{r}\right)^{2}n_{sf}\left(\boldsymbol{r}\right)$
is ``carved out'' of $n_{sf}\left(\boldsymbol{r}\right)$ using
a carving function $c_{f}\left(\boldsymbol{r}\right)^{2}$. Thus:
\begin{equation}
n_{f}\left(\boldsymbol{r}\right)=c_{f}\left(\boldsymbol{r}\right)^{2}\sum_{n}\left(p_{n}^{f}\right)^{2}\psi_{n}^{f}\left(\boldsymbol{r}\right)^{2},\label{eq:carved-density}
\end{equation}
 where, inspired by Hirshfeld partitioning,~\cite{Hirshfeld1977a}
the carving function is defined as: 

\[
c_{f}(\boldsymbol{r})=\sqrt{\frac{\sum_{a\in f}n_{a}^{\left(0\right)}\left(\boldsymbol{r}\right)}{\sum_{a\in sf}n_{a}^{\left(0\right)}\left(\boldsymbol{r}\right)}},
\]
where $n_{a}^{\left(0\right)}\left(\boldsymbol{r}\right)$ is the
spherical density of neutral atom $a$. The temperature parameter
$\beta$ in the definition of the population $p_{n}^{f}$ is chosen
be the same value as that of the sDFT calculation, while the chemical
potential $\mu_{f}$ of each fragment is determined by the condition
of neutrality of the fragment: 
\begin{equation}
\int n_{f}\left(\boldsymbol{r}\right)d\boldsymbol{r}=\int\sum_{a\in f}n_{a}^{\left(0\right)}\left(\boldsymbol{r}\right)d\boldsymbol{r}.\label{eq:neutrality}
\end{equation}
Defining non-orthogonal functions $\tilde{\psi}_{n}^{f}\left(\boldsymbol{r}\right)=c_{f}\left(\boldsymbol{r}\right)p_{n}^{f}\psi_{n}^{f}\left(\boldsymbol{r}\right)$,
the fragment density of Eq.~(\ref{eq:carved-density}) becomes $n_{f}\left(\boldsymbol{r}\right)=2\sum_{n}\tilde{\psi}_{n}^{f}\left(\boldsymbol{r}\right)^{2}$,
so the chemical potential is determined from the condition: 
\begin{equation}
2\sum_{n}\left\langle \tilde{\psi}_{n}^{f}\left|\tilde{\psi}_{n}^{f}\right.\right\rangle =\int\sum_{a\in f}n_{a}^{\left(0\right)}\left(\boldsymbol{r}\right)d\boldsymbol{r}.
\end{equation}

After determining $\mu_{f}$ and in order to construct the reduced
density matrix (RDM), we orthogonalize the functions $\tilde{\psi}_{n}^{f}\left(\boldsymbol{r}\right)$
by diagonalizing the overlap matrix $S_{nn^{\prime}}^{f}=\left\langle \tilde{\psi}_{n}^{f}\left|\tilde{\psi}_{n^{\prime}}^{f}\right.\right\rangle $,
obtaining the unitary matrix $U_{f}$ of eigenvectors and the eigenvalues
$s_{n}^{f}>0$ (so that $U_{f}^{T}S_{f}U_{f}=diag\left[s_{1}^{f},s_{2}^{f}\dots\right]$).
The orthogonal wavefunctions are: $\phi_{m}^{f}\left(r\right)=\sum_{n}\tilde{\psi}_{n}^{f}\left(r\right)U_{nm}^{f}$
and the norm is $\left\langle \phi_{m}^{f}\left|\phi_{m}^{f}\right.\right\rangle =s_{m}^{f}$.
Using the new wave functions, the unsaturated fragment density is
given by:
\[
n_{f}\left(\boldsymbol{r}\right)=2\sum_{m}\phi_{m}^{f}\left(\boldsymbol{r}\right)^{2}
\]
and the RDM by 
\[
\hat{\theta}_{f}=2\sum_{m}\left|\phi_{m}^{f}\right\rangle \left\langle \phi_{m}^{f}\right|.
\]
Using the RDM we express the unsaturated fragment expectation value
appearing in Eq.~(\ref{eq:fragment-correction-A}) as:

\[
\left\langle \hat{A}_{f}\right\rangle \equiv tr\left[\hat{\theta}_{f}\hat{A}\right]=tr\left[\sqrt{\hat{\theta}_{f}}\hat{A}\sqrt{\hat{\theta}_{f}}\right],
\]
 where
\begin{equation}
\sqrt{\hat{\theta}_{f}}=\sqrt{2}\sum_{m}\left(s_{m}^{f}\right)^{-1/2}\left|\phi_{m}^{f}\right\rangle \left\langle \phi_{m}^{f}\right|.
\end{equation}
By choosing the fragment grid-points to be a subset of the full system
grid, each stochastic orbital $\chi_{i}$ ($i=1,\dots,I$) of the
full system appears as a stochastic orbital on the fragment grid and
can be used to perform the stochastic estimate appearing in Eq.~(\ref{eq:fragment-correction-A})
as:
\[
\left\langle \hat{A}_{f}\right\rangle ^{I}=\frac{1}{I}\sum_{i}\left\langle \chi_{i}\left|\sqrt{\hat{\theta}_{f}}\hat{A}\sqrt{\hat{\theta}_{f}}\right|\chi_{i}\right\rangle _{f},
\]
where the subscript $f$ on the left denotes integration over the
fragment grid. The difference $\Delta A_{f}^{I}=\left\langle \hat{A}_{f}\right\rangle -\left\langle \hat{A}_{f}\right\rangle ^{I}$
in Eq.~(\ref{eq:fragment-correction-A}) can now be written in a
unified form as:
\begin{equation}
\Delta A_{f}^{I}=2\sum_{mm^{\prime}}\Delta_{mm^{\prime}}^{f\,I}\left\langle \phi_{m}^{f}\left|\hat{A}\right|\phi_{m^{\prime}}^{f}\right\rangle _{f},\label{eq:DeltaA-I}
\end{equation}
where:

\begin{equation}
\Delta_{mm^{\prime}}^{f\,I}\equiv\delta_{mm^{\prime}}-\frac{1}{I}\sum_{i}\frac{\left\langle \chi_{i}\left|\phi_{m}^{f}\right.\right\rangle _{f}\left\langle \phi_{m^{\prime}}^{f}\left|\chi_{i}\right.\right\rangle _{f}}{\sqrt{s_{m}^{f}s_{m^{\prime}}^{f}}}.
\end{equation}
Hence, by calculating the matrix $\Delta_{mm^{\prime}}^{f\,I}$ all
types of expectation value corrections can be obtained from Eq.~(\ref{eq:DeltaA-I}).

The efficacy of embedded fragments in sDFT force calculations is achieved
through a reduction of the STD $\sigma\left(f_{x}\right)$ of a force
component. The STD $\sigma\left(f_{x}\right)$ is proportional to
$1/\sqrt{I}$, where $I$ is the number of stochastic orbitals and
the proportionality constant, denoted $\sigma_{1}\left(f_{x}\right)=\sqrt{I}\sigma\left(f_{x}\right)$,
is called the \emph{inherent }STD. This quantity depends on the NC
characteristics but not on the number of stochastic orbitals. In Fig.~\ref{fig:std0-Frags}
we plot the inherent STD on each atom for $\text{Si}_{35}\text{H}_{36}$
and $\text{Si}_{705}\text{H}_{300}$ as a function of fragment size.
Even the use of the smallest fragments reduces the inherent force
STD by a significant factor, $1.6$ (for $\text{Si}_{705}\text{H}_{300}$)
to $2.3$ (for $\text{Si}_{35}\text{H}_{36}$). Using larger fragments
reduces the STD by an additional factor of $\approx1.5$, with increasing
effect for larger systems, since the electron density in the larger
fragments is similar to that of the full system. It is interesting
to see that for the forces there is no noticeable sublinear scaling:
the inherent STD for both systems is similar, with the larger system
having a slightly ($\approx5\%$) STD. 

In summary, the embedded fragment sDFT method serves as a way to expedite
the sDFT calculation by a judicious choice of fragment size and composition.
As the fragment size grows, the numerical effort invested in sDFT
decreases (due to reduction of STD) while in dDFT it increases. For
example, consider Fig.~\ref{fig:std0-Frags} where we showed that
increasing the fragment size by a factor of $10-20$ reduces the STD
by a factor of $2$ and therefore the sDFT CPU time by a factor of
$\approx2^{2}=4$. On the other hand since the fragments are ten-fold
larger, the amount of dDFT work on them increases (cubically) by a
factor of more than$\sim10^{3}$. Clearly then, the optimal fragment
size is system dependent. Embedded fragments have the additional benefit
of providing an initial density for the SCF calculation, significantly
reducing the number of SCF cycles.

\section{\label{sec:Hellmann-Feynman-Forces}Stochastic estimates of the forces
and energy perturbations }

In the stochastic method, the electronic density is (see Eq.~(\ref{eq:stoch-dens})):
\begin{equation}
n\left(r\right)=2\left\langle \chi\left|\hat{P}\delta\left(r-\hat{r}\right)\hat{P}\right|\chi\right\rangle 
\end{equation}
where $\hat{P}=\sqrt{\theta_{\mu}}$ is the Chebyshev expansion of
the projection operator, depending on $\beta$ and $\mu$, on the
occupied space of $\hat{h}_{KS}$, and the energy is 

\begin{align}
E & =2\left\langle \chi\left|\hat{P}\hat{T}\hat{P}\right|\chi\right\rangle +\int v_{eN}\left(r;R\right)n\left(r\right)dr+E_{HXC}\left[n\right]\\
 & =2\left\langle \chi\left|\hat{P}\left[\hat{T}+v_{eN}\left(r;R\right)\right]\hat{P}\right|\chi\right\rangle +E_{HXC}\left[n\right]\nonumber 
\end{align}
where $E_{HXC}\left[n\right]$ is the Hartree-exchange-correlation
energy functional, depending only on the electronic density $n\left(\boldsymbol{r}\right)$.
Under variation in position of nuclei $R$:
\begin{align}
\delta E & =2\left\langle \chi\left|\hat{P}\left[\hat{T}+v\left(\hat{r},R\right)\right]\delta\hat{P}\right|\chi\right\rangle \\
 & +2\left\langle \chi\left|\delta\hat{P}\left[\hat{T}+v\left(\hat{r},R\right)\right]\hat{P}\right|\chi\right\rangle \nonumber \\
 & +\left\langle \chi\left|\hat{P}\delta v\left(\hat{r},R\right)\hat{P}\right|\chi\right\rangle \nonumber \\
 & +\int v_{HXC}\left(r\right)\delta n\left(r\right)dr\nonumber 
\end{align}
which using 
\begin{align}
\delta n\left(r\right) & =2\left\langle \chi\left|\delta\hat{P}\delta\left(r-\hat{r}\right)\hat{P}\right|\chi\right\rangle \\
 & +2\left\langle \chi\left|\hat{P}\delta\left(r-\hat{r}\right)\delta\hat{P}\right|\chi\right\rangle \nonumber 
\end{align}
can be written as:
\begin{align}
\delta E & =2\left\langle \chi\left|\hat{h}_{KS}\hat{P}\delta\hat{P}+\delta\hat{P}\hat{P}\hat{h}_{KS}\right|\chi\right\rangle \\
 & +\left\langle \chi\left|\hat{P}\delta v\left(\hat{r},R\right)\hat{P}\right|\chi\right\rangle \nonumber \\
 & =2\left\langle \chi\left|\hat{h}_{KS}\hat{P}\delta\hat{P}+\delta\hat{P}\hat{P}\hat{h}_{KS}\right|\chi\right\rangle \nonumber \\
 & +\int n\left(\boldsymbol{r}\right)\delta v\left(\boldsymbol{r};R\right)d^{3}r\nonumber 
\end{align}

The average of the second term on the right leads to the work of the
Hellmann-Feynman force Eq.~(\ref{eq:hellman-Fey Force}) while the
first term can be shown to vanish when a full sampling is made on
$\chi$ and when $\beta\to\infty$for then $P\delta PP=0$. 

\bibliographystyle{aipnum4-1}
\bibliography{RoiBaerLib}

\end{document}